\documentclass[openacc]{rstransa}  

\begin{document}

\title{Controlling crystallization: What liquid structure and dynamics reveal about crystal nucleation mechanisms}

\author{Jutta Rogal$^{1,2}$ and Grisell D{\'i}az Leines$^{3}$}

\address{
$^{1}$Department of Chemistry, New York University, New York, NY 10003, USA\\
$^{2}$Fachbereich Physik, Freie Universit{\"a}t Berlin, 14195 Berlin, Germany \\
$^{3}$Yusuf Hamied Department of Chemistry, University of Cambridge, Lensfield Road, Cambridge CB2 1EW, United Kingdom
}

\subject{computational physcics, computational chemistry, chemical physics, physical chemistry, material science}


\corres{Jutta Rogal\\
\email{jutta.rogal@nyu.edu}}

\begin{abstract}
Over recent years, molecular simulations have provided invaluable insights into the microscopic processes governing the initial stages of crystal nucleation and growth.
A key aspect that has been observed in many different systems is the formation of precursors in the supercooled liquid that precedes the emergence of crystalline nuclei. The structural and dynamical properties of these precursors determine to a large extend the nucleation probability as well as the formation of specific polymorphs.  
This novel microscopic view on nucleation mechanisms has further implications for our understanding of the nucleating ability and polymorph selectivity of nucleating agents, as  these appear to be strongly linked to their ability in modifying structural and dynamical characteristics of the supercooled liquid, namely liquid heterogeneity.
In this perspective, we highlight recent progress in exploring the connection between liquid heterogeneity and crystallization, including the effects of templates, and the potential impact for controlling crystallization processes.

\end{abstract}



\begin{fmtext}

\end{fmtext}

\maketitle

\section{Introduction}

Liquid-solid phase transitions are physical processes ubiquitous in nature. They are essential in biological processes, pharmaceutical applications, materials design, and environmental research, among others. Yet, a thorough fundamental understanding of crystallization remains elusive. Crystallization often starts with a nucleation event through the contact with a surface or impurities. The nucleation step determines to a large 
extent
the crystallization pathway at the very early stages of the process, which often proves to be a complex microscopic event, involving an intricate interplay between dynamical and structural physical descriptors in the liquid and at the surface~\cite{Sosso2016,Jungblut2016}. Even with combined efforts of the latest experimental techniques and computational modelling approaches, quantitative predictions of crystal nucleation mechanisms have remained bottlelnecked by the time scales and  microscopic complexity that govern crystal nucleation~\cite{Sosso2016,Blow2021}. 

In the last decades, classical nucleation theory (CNT)~\cite{Becker1935,Binder1987} has provided a successful phenomenological description of crystal nucleation, but often yields quantitative inconsistencies ~\cite{Sosso2016,Blow2021}. This is thought to be attributable to the simplifying assumptions of CNT: a spherical shape of the nucleus, liquid homogeneity with random fluctuations, and small clusters sharing the same thermodynamic phase as the bulk with a sharp liquid-solid interface. But even for homogeneous systems and simple liquids it is well known that crystallization pathways are often far from the classical scenario~\cite{Jungblut2016,Gebauer2014,tenWolde1997,tenWolde1999,Zhang2007,prestipino_SystematicImprovement_2012,Russo2012, Sear2012, DLeines2017,Lechner2010}. For heterogeneous nucleation, other factors, such as template morphology, absorption, and the local ordering of the contact liquid layer can  impact  nucleation mechanisms, and the classical scenario cannot be confirmed even for simple model interfaces~\cite{fitzner_PredictingHeterogeneous_2020, PhysRevLett.108.025502,Jungblut2013,Page2009,doi:10.1063/1.4961652,PhysRevLett.128.166001}. 

Recent progress in advanced computational techniques, including enhanced sampling methods and artificial intelligence approaches, 
facilitates
to model crystallization more precisely than ever, hinting an increase in the accuracy and predictive knowledge of complex processes~\cite{DLeines2018,fitzner_PredictingHeterogeneous_2020,Lechner2010,Piaggi_icenuc}. Indeed, making use of these methodologies, it has been established in recent studies that liquids are not homogeneous as proposed by CNT, but reveal hidden structural order and collective dynamical behaviour occurring within fluctuations. This phenomenon, referred to as dynamical and structural heterogeneity in the liquid, has been found to play a dominant role in crystal nucleation and disclose information about the polymorphic outcomes of crystallization~\cite{Gebauer2014,tenWolde1997,tenWolde1999,Zhang2007,prestipino_SystematicImprovement_2012,Russo2012}. Consequently, crystallization does not start from random fluctuations in the liquid but with the formation of precursors, i.e.  pre-ordered regions that exhibit either increased bond-orientational order, density, or reduced mobility. 
These precursors facilitate the formation of crystal nuclei, presumably by decreasing the crystal-liquid interfacial free energy~\cite{DLeines2018,Tanaka2012, PhysRevX.8.021040}, and signal the polymorphs that will be selected~\cite{Gebauer2014,tenWolde1997,tenWolde1999,Zhang2007,prestipino_SystematicImprovement_2012,Russo2012}. Several studies on crystal nucleation, including ice formation~\cite{fitzner_IceBorn_2019}, crytallization in metals~\cite{DLeines2017,DLeines2018,Zhang2019,doi:10.1063/5.0017575,Hu2022}, 
hard spheres~\cite{PhysRevLett.105.025701,PhysRevLett.96.175701}, and colloidal models~\cite{Lechner2011a,Tan2014}, among others~\cite{Sear2012} confirm a correlation between liquid heterogeneity and nucleation events, as well as polymorph selection. 

 In view of these findings, an interesting question is how interfaces modify the liquid structure and dynamics in connection to crystallization. Far from being only an interesting correction to CNT, if liquid heterogeneity at interfaces provides fingerprints for the nucleating ability of a surface and the crystallization pathway, a promising alternative to acquire fast and predictive knowledge of these processes and screen interfaces is at hand. With the rapid development of deep learning techniques for molecular simulations and the increasing availability of large data sets, the ambition of classifying  the nucleating ability of surfaces based on the properties of the liquid can be accomplished. First promising ideas to predict the nucleating ability of model interfaces during ice formation have been recently introduced~\cite{davies_AccuratePrediction_2022}. Nevertheless, an imminent need of fundamental understanding of crystal nucleation processes is still essential in providing the relevant microscopic factors and physical-chemical descriptors that  serve as inputs for training machine learning methods and meaningful screening of the nucleating ability of interfaces and materials. To this end, enhanced sampling simulations of nucleation processes~\cite{DLeines2018,PhysRevLett.128.166001} together with deep learning approaches are making it possible to access extended timescales and even use electron interactions to model crystallization in larger systems~\cite{Piaggi_icenuc}.
 
 In this article we present an overview of recent advances and future perspectives on
 some
 non-classical crystal nucleation mechanisms and what they reveal about the role of fluctuations and liquid heterogeneity in the crystallization of supercooled liquids.

\section{Precursor-mediated mechanisms}
\label{sec:homo_prec}

It is well established that structural and dynamical heterogeneities can be present in supercooled liquids. Yet, only in recent years we have witnessed increasing evidence suggesting that liquid heterogeneity  plays a crucial role in crystallization mechanisms and in determining the glass formation ability of various materials. One of the main assumptions in CNT~\cite{Becker1935,Binder1987} is that crystallization events are driven by random fluctuations in the homogeneous liquid. Nevertheless, the formation of precursor regions in the liquid highlights the crucial role of thermal fluctuations and heterogeneity in driving crystallization events. Indeed,  precursor-mediated nucleation mechanisms during crystallization, namely 'two-step' mechanisms, have been reported in several studies and for various materials, including colloids, hard spheres, ice, metal alloys, and proteins~\cite{PhysRevLett.105.025701,Russo2012,DLeines2017,doi:10.1063/5.0017575,Lechner2011a,Tan2014,Zhang2019,Sosso2016}. These mechanisms are characterized by the initial formation of pre-ordered regions in the liquid that precede the formation of crystallites within the centres of the precursors. 
The structural characteristics of these regions also pre-determine the preferred polymorphic structure that will grow. 
The precursors exhibit either increased bond-orientational order~\cite{Russo2012,DLeines2017,doi:10.1063/5.0017575}, density~\cite{PhysRevLett.96.046102,doi:10.1021/cg049977w,tenWolde1997,tenWolde1999} or mobility~\cite{fitzner_IceBorn_2019,diazleines_InterplayStructural_2022} which promotes the emergence of crystallites, presumably by reducing the interfacial free energy. 
    
The formation of high-density aggregates in the liquid that precede crystal nucleation was first identified in Refs.~\cite{tenWolde1997,tenWolde1999,PhysRevLett.96.046102}.  In these studies, 
the coupling between a critical concentration of density fluctuations and translational ordering during crystallization was revealed to play a key role in nucleation. 
Other experimental and theoretical studies have shown the importance of density fluctuations in the formation of precursors~\cite{PhysRevLett.96.175701,PhysRevLett.102.198302,doi:10.1063/1.1992475} and two-step mechanisms. Recent works have demonstrated the significance of another kind order fluctuations, namely regions of higher bond-orientational order in the liquid, that act as precursors of crystallization. Tanaka {\it et al.}~\cite{Russo2012,Kawasaki2011} found that crystallization is not described by translational ordering of a density field, but by directional bonding in colloidal systems and hard spheres. This behaviour was explained by a weak coupling between density  and bond-order fluctuations. Moreover, the precursors were found to exhibit symmetries that resemble the  crystal structure that grows during crystallization, highlighting the important role of precursors during polymorph selection. 
    
\begin{figure}
\centering\includegraphics[width=\textwidth]{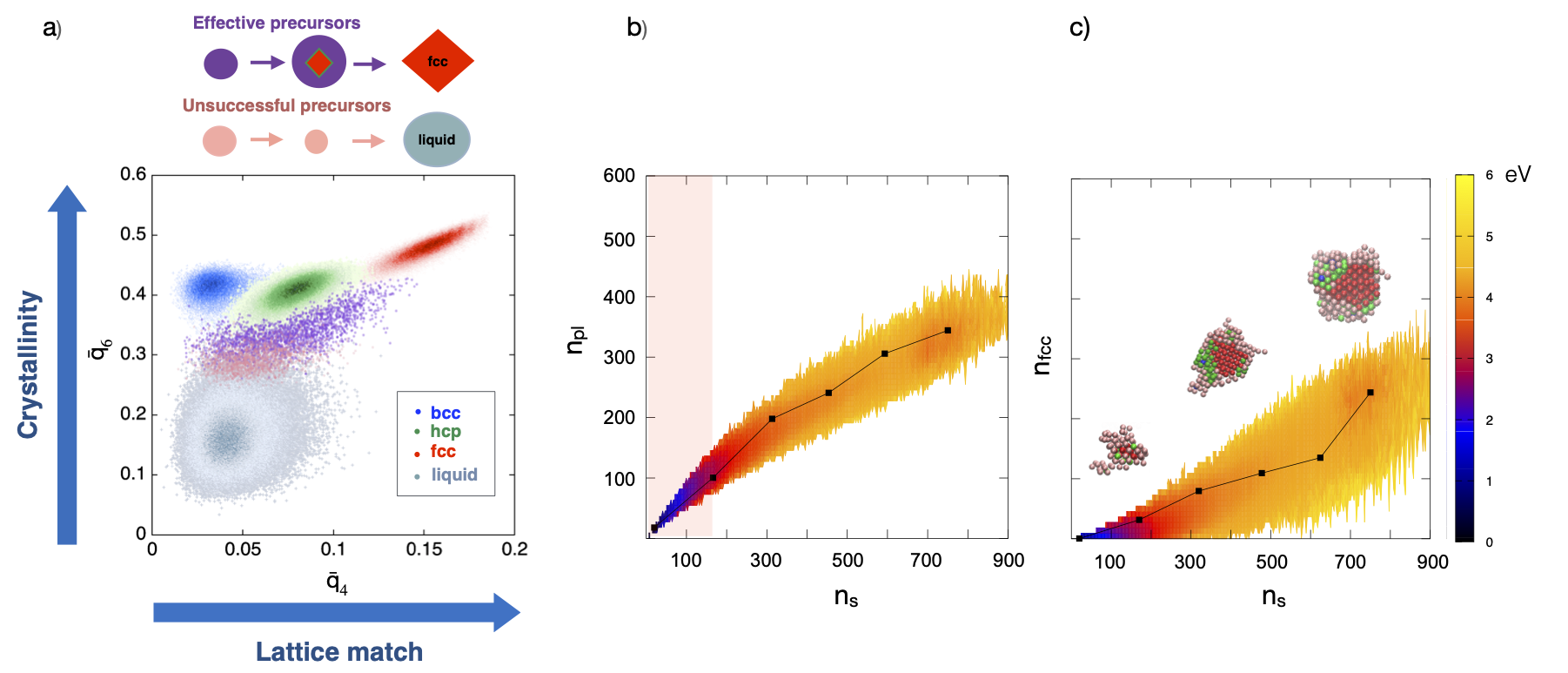}
\caption{
\label{fig:Ni-homo-nuc}
Precursor-induced mechanism of homogeneous nucleation in nickel. 
The local environment around each atom was characterised by the averaged local bond-order parameters $\bar{q}_4$ and  $\bar{q}_6$~\cite{Lechner2008}.
(a) $\bar{q}_4-\bar{q}_6$ reference distributions for crystal structures in nickel. The pre-ordered liquid is a mesocrystal phase that lies between liquid and crystal symmetries (pink and purple dots). The $\bar{q}_4-\bar{q}_6$ values for pre-ordered liquid particles in pre-critical clusters that yield nucleation events (purple dots) show that crystal precursors exhibit enhanced crystallinity (bond-orientational order). 
In contrast, pre-critical clusters  
that dissolve to the liquid phase (pink dots) are characterized by a reduced crystallinity. 
(b) and (c) Average crystallization paths on 2D free energy projections with the number of (b) pre-structured particles  $n_\text{pl}$  and (c) the number of fcc particles $n_\text{fcc}$ in the largest cluster vs. the size of the largest cluster, $n_s$. In the early stages of crystallization,  
$n_\text{pl}$ increases linearly as a function of 
$n_s$ (pink transparent region in (b)), indicating the initial formation of precursors that precede the emergence of crystallites. In contrast, 
as $n_s$ increases,  $n_\text{fcc}$ is negligible 
for cluster sizes $< 150$ particles,
corresponding to $\sim 1$~eV in the energy barrier. Reprinted with permission from G. D\'{i}az Leines and J. Rogal, J. Phys. Chem. B 122, 10934-10942 (2018)~\cite{DLeines2018}. Copyright 2018 American Chemical Society. Reprinted figure with permission from G. D{\'i}az Leines and J. Rogal,  Phys. Rev. Lett. 128 (16), 166001 (2022)~\cite{PhysRevLett.128.166001}. Copyright 2022 by the American Physical Society.}
\end{figure}

Previously, we investigated other precursor-mediated mechanisms during crystal nucleation in metals, such as nickel~\cite{DLeines2017,DLeines2018} and molybdenum~\cite{doi:10.1063/5.0017575}, using transition interface sampling (TIS) simulations~\cite{Dellago2002,VanErp2005,Rogal2010}. In these systems, the initial emergence of pre-ordered liquid regions with higher bond-orientational order plays a crucial role in the structural description of the growing nucleus and its interfacial free energy. In agreement with Tanaka's observations, we find that these mesocrystal clusters act as precursors or seeds for crytallites that grow embedded within their centres (see Fig.~\ref{fig:Ni-homo-nuc}). Using a maximum likelihood analysis of the path ensemble data~\cite{DLeines2018}, we demonstrated quantitatively that taking into account the pre-ordering in the liquid significantly  
improves the description of the nucleation mechanism.
This evidences that crystal precursors are not  trivial fluctuations of order that precede translational order but an essential step in the nucleation process.
Moreover, the precursors were found to exhibit structural symmetries that resemble the emerging crystalline  phase, establishing a link between structural heterogeneity in the liquid and polymorph selection.  

Another example is ice nucleation. Fitzner {\it et al.}~\cite{fitzner_IceBorn_2019} employed transition path sampling (TPS) and MD simulations to investigate homogeneous nucleation in ice. 
The crystallization paths showed that precursor regions in the liquid are characterized by ice-like structures with an abundance of 6-membered hydrogen-bonded rings, which are structural hallmarks of the nucleating ice. 
They also linked the formation of structural precursors to the dynamical properties of the liquid, which will be discussed in section~\ref{sec:dynhet}.
    
Russo {\it et al.}~\cite{PhysRevX.8.021040} recently showed that, generally, the structural features of liquids can control the glass-forming and crystal-forming ability of a system, by suppressing or promoting the formation of precursors via a thermodynamic interface penalty.  When crystal-like angular order is present in supercooled liquids, there is weak frustration against crystallisation.  But the inhibition of fluctuations of crystal-like order in liquids signals frustration and increased amorphization ability of a system.
    
So far, precursor-mediated mechanisms have been confirmed in a variety of simple and more complex liquids, highlighting the importance of structural fluctuations and liquid heterogeneity during crystallization and  polymorph selection. In this scenario, it becomes highly desirable to explore how targeted modifications and manipulations of the structural features in liquids can impact the crsytallization pathways and signal the nucleating ability of templates and materials. In section~\ref{sec:heteronuc}, we will discuss recent advances in our understanding of  `non-classical'  heterogeneous nucleation mechanisms and its consequences in material design.

\section{Dynamical heterogeneity}
\label{sec:dynhet}
In addition to structural pre-ordering, the dynamical properties of supercooled liquids can likewise play a prominent role in the nucleation process.  Dynamical heterogeneity (DH) is mainly discussed in the context of glass formation~\cite{berthier_TheoreticalPerspective_2011} and refers to spatiotemporal fluctuations of the dynamics.  
These fluctuations are spatially extended and form distinct regions with different mobilities.
The self-intermediate scattering function provides information concerning the relaxation dynamics 
as a function of the reciprocal space vector $\mathbf{q}$  which, in  isotropic systems, can be averaged over independent directions, and exhibits characteristic features for a given length scale $q = || \mathbf{q} || $ with increasing DH.  
In particular, a plateau region evolves that is associated with particles that are confined in their movement by their neighbours. 
The dynamic susceptibility, which can be defined as the variance of the self-intermediate scattering function~\cite{toninelli_DynamicalSusceptibility_2005a}, initially increases with time as the dynamics heterogeneity slowly builds up and exhibits a peak  at a characteristic time $t^{*}$ after which it decrease back to zero as time goes to infinity~\cite{berthier_TheoreticalPerspective_2011}. 
This time $t^{*}$ can also be considered as the time of maximum heterogeneity.

The intermediate scattering function and dynamical susceptibility describe the average DH of the bulk liquid.  In simulations, the relative mobility of individual particles $i$ can be spatially resolved by computing the dynamical propensity (DP)~\cite{sosso_DynamicalHeterogeneity_2014,fitzner_IceBorn_2019}
\begin{equation}
\label{eq:dp}
\text{DP}_i (t^{*}) = \left\langle\frac{|| \mathbf{r}_i(t^{*}) - \mathbf{r}_i(0) ||^2}{\text{MSD}}\right\rangle_{\text{ISO}} \quad ,
\end{equation}
where MSD is the the mean-squared displacement at time $t^{*}$ and $\langle \dots \rangle_{\text{ISO}}$ is the average over an isoconfigurational ensemble~\cite{widmer-cooper_StudyCollective_2007,colombo_MicroscopicPicture_2013}.    
It can be computed by running a number of MD trajectories for a given configuration with random Maxwell-Boltzmann distributed velocities.  The relative mobility of a particle depends on the chosen length scale $q^{*}$ and time scale $t^{*}$.  Setting $q^{*}$ to the first peak of the structure factor and $t^{*}$ to the peak in the dynamical susceptibility characterises a regime where the dynamics in the nearest-neighbour environment is most heterogeneous.  The computational effort depends on $t^{*}$ and, in general, several representative configurations are needed to obtain distributions of DP values, which makes the evaluation rather costly.  The central advantage, however, is that the spatially resolved mobility can directly be correlated with structural features and the nucleation processes, correspondingly.

The DH of supercooled liquids strongly depends on the temperature and is more pronounced in glass forming materials but was shown to impact nucleation even for seemingly homogeneous systems.  Here, we discuss three examples with very different degrees of DH.
GeTe has been investigated as a prototypical phase change material exhibiting very fast crystallisation kinetics.    In~\cite{sosso_DynamicalHeterogeneity_2014}, the high crystallisation speed was linked to the formation of Ge-Ge chains that modify the atomic mobility based on MD simulations. 
Close to the glass transition temperature, GeTe exhibits significant DH. 
Specifically, regions of high and low mobility were identified showing clustering of atoms with large (most mobile, MM) and small (most immobile, MI) DP values.  The Ge-Ge chains are located in the MM regions and it was suggested that the enhanced mobility in spatially confined regions close to crystalline nuclei speeds up the crystallisation kinetics.  The crystallisation mechanisms is thus affected by the interplay between the structural and dynamical features of the liquid.

Similarly, the connection between mobility and the nucleation of crystalline clusters in supercooled water has been investigated with MD and TPS~\cite{fitzner_IceBorn_2019}.
A strong spatial correlation was found between regions of MI molecules and the formation of pre-critical (ice-like) clusters.  Furthermore, the decrease in mobility appeared to precede the occurrence of structural ordering which can be viewed as a {\it dynamical incubation period}. The analysis of ice-/liquid-like and mobile/immobile regions along nucleation trajectories from TPS simulations revealed that the crystalline ice clusters indeed emerge and grow within  low mobility regions.
The relatively mobile and immobile domains also exhibited distinct structural features.  Specifically, the MI regions contain a large number of $6 \pm 1$ membered hydrogen-bonded rings that resemble ice-like structural motifs.  This correlation between decreased mobility and crystalline features was seen as the link between immobile domains and nucleation.

\begin{figure}
\centering\includegraphics[width=\textwidth]{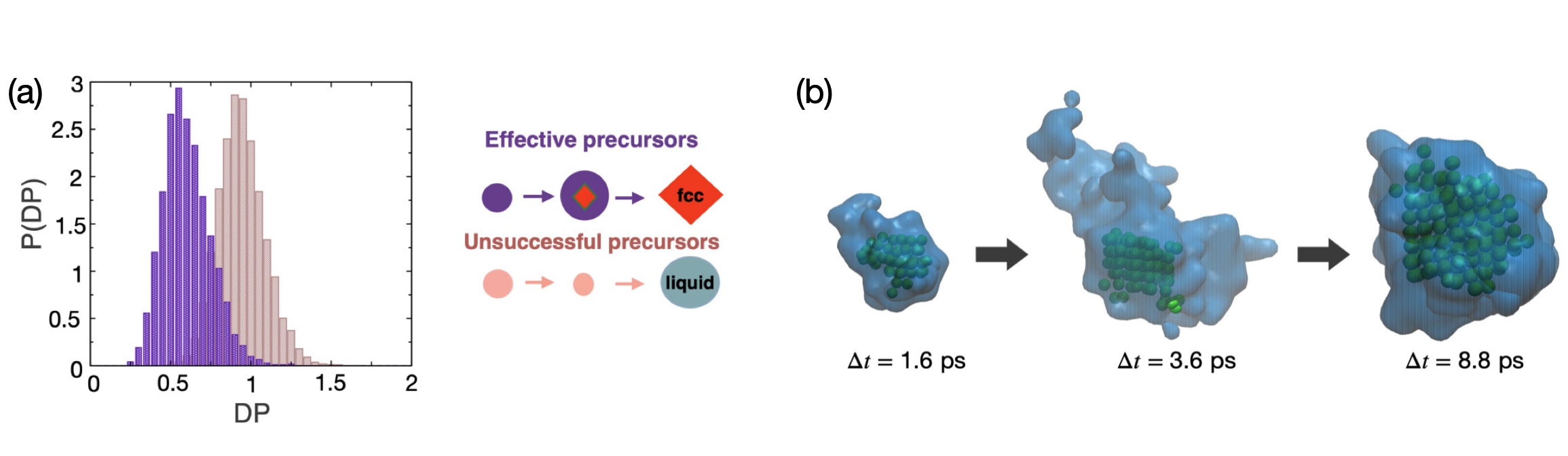}
\caption{
\label{fig:Ni-DH}
(a) Probability distribution of dynamical propensity values for atoms in effective (purple) and unsuccessful (pink) precursors.  The distribution for effective precursors shows a clear shift to lower mobilities. 
(b) Configurations along a nucleation trajectory from the TPS ensemble. The blue transparent region marks the largest cluster of most immobile atoms, the green spheres indicate the largest solid cluster.  The time $\Delta t$ corresponds to the transition time after leaving the stable liquid state in the TPS simulation.
Reproduced from Ref.~\cite{diazleines_InterplayStructural_2022} with permission from the Royal Society of Chemistry.}
\end{figure}
Another interesting example is homogeneous nucleation in Ni already discussed in Section~\ref{sec:homo_prec}.
Supercooled liquid Ni does not exhibit a large degree of DH when analyzing the intermediate scattering function and probability distribution of DP values over the entire bulk volume.  This is also expected as elemental Ni is a poor glass former.  
During nucleation, however, the local formation of immobile regions plays a key role associated with the emergence of dynamical precursors~\cite{diazleines_InterplayStructural_2022}.
From our TPS simulations, we analysed trajectories where structural precursors of size $n_s = 50$ particles either continued to grow and crystallised or dissolved again.  The structural features of these {\it effective} and {\it unsuccessful} precursors are similar in both cases and clearly distinct from the liquid as well as from the crystalline phases.  The dynamical properties of the precursors are, however, quite different.  The distribution of DP values for unsuccessful precursor is similar to that of the supercooled liquid, whereas the DP distribution of effective precursors shows a noticeable shift to lower mobilities, shown in Fig.~\ref{fig:Ni-DH}(a).  It appears that structural pre-ordering alone without a decrease in mobility, as seen for unsuccessful precursors, does not promote the emergence and subsequent growth  of crystalline nuclei.
Similar to the observations for water, domains of low mobility appear first in the supercooled liquid followed by the formation of solid-like structural precursors and, eventually, the emergence of crystalline clusters.  
Such an analysis required the computation of DP values for every configuration along the nucleation trajectories in the TPS ensemble together with a clustering of atoms with low DP values. 
As the solid cluster continues to grow, it remains surrounded by a region with decreased mobility as shown in Fig.~\ref{fig:Ni-DH}(b), which is also of interest in the context of growth dynamics.
It should be noted that there is no correlation between low mobility regions and local temperature fluctuations.  The reduced mobility appears to originate from structural features that interfere with the movement of atoms. 

The strong spatial and temporal correlation between the formation of low mobility regions and structural precursors in the supercooled liquid followed by the emergence of the crystalline phase has been observed for systems with very different degrees of DH.  This suggests a general mechanisms for nucleation where the initial step is characterised by the appearance of domains with reduced mobility.

\section{Template induced precursor formation}
\label{sec:heteronuc}

Crystallization typically initiates through the contact of a liquid layer with an interface or impurity, which enhances crystal nucleation by reducing the (homogeneous) free energy barrier of the process. Thus, heterogeneous nucleation dominates in nature and is key for most crystallization processes. The classical views of heterogeneous nucleation generally establish that the nucleating ability of a template is determined mostly by the degree of lattice matching or how commensurate  the symmetry and density of the template is with the crystal phase of the growing solid cluster~\cite{Turnbull1952}. However, it is well known that the scenario is much more complicated and often other factors, such as template morphology, absorption and local ordering of the contact liquid layer can modify the crystallization paths dramatically~\cite{fitzner_PredictingHeterogeneous_2020, PhysRevLett.108.025502,Jungblut2013,Page2009,doi:10.1063/1.4961652}.  Major gaps remain in our understanding of which factors determine the nucleating ability of interfaces or impurities.

In light of the recent strong evidence of an interconnection between liquid heterogeneity and crystallization, there is an imminent need to understand what the hidden local structural and dynamical features of liquids can disclose about the nucleating ability of materials and if they are key in the selective rules that determine crystallization pathways. The importance of pre-critical fluctuations in the mechanism of heterogeneous nucleation was initially discussed by Fitzner {\it et al.}~\cite{fitzner_precrfluc_2017} in a study of ice nucleation at model interfaces using MD simulations. In this work, it was observed that, contrary to the assumptions of CNT about random fluctuations in the liquid, pre-critical fluctuations can reveal different structural features at interfaces that have the same nucleation temperature. This result suggests that fluctuations of order in water layers in contact with an interface connote the polymorph that will be selected during freezing.   

\begin{figure}
\centering\includegraphics[width=\textwidth]{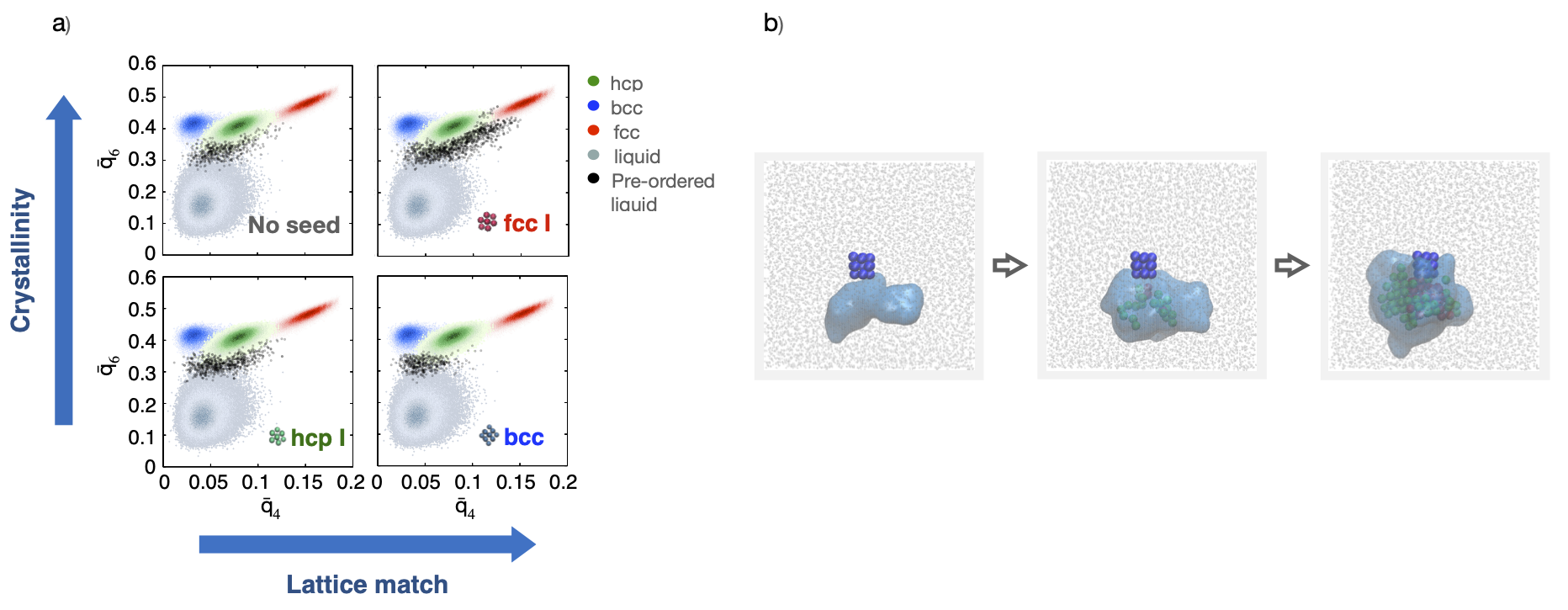}
\caption{
Template induced precursor formation in heterogeneous nucleation in nickel.
The local environment around each atom was characterised by the averaged local bond-order parameters $\bar{q}_4$ and  $\bar{q}_6$~\cite{Lechner2008}.
(a)  $\bar{q}_4-\bar{q}_6$ reference maps for crystal structure identification (fcc, bcc, hcp, and liquid). The crystallinity of   pre-ordered liquid particles (black) during liquid fluctuations  in the presence of various seeds. A clear increase in crystallinity (bond orientational order) is observed for the fcc seed, indicating that the seeds modify the bond orientational order of the liquid during pre-critical fluctuations according to their efficiency to promote nucleation events.  (b) Snapshots of representative nucleation trajectory in the presence of an fcc seed. The pre-structured liquid region (blue transparent surface) forms initially near the seed and precedes the emergence of the crystalline cluster within the center of the precursor. The crystalline cluster is  composed of fcc (red atoms) and random-hcp (green atoms). The fcc seed (blue atoms) is  surrounded predominantly by pre-structured liquid and random-hcp and mostly positioned at the surface of the nucleus. Reprinted figure with permission from G. D{\'i}az Leines and J. Rogal, Phys. Rev. Lett. 128 (16), 166001 (2022)~\cite{PhysRevLett.128.166001}. Copyright 2022 by the American Physical Society.
\label{fig:Ni-hetnuc}
}
\end{figure}

Recently, we have performed an extensive study of crystal nucleation in nickel in the presence of small seeds, using TIS simulations~\cite{PhysRevLett.128.166001}. 
The structural heterogeneity in the supercooled liquid was directly linked to the crystal nucleation mechanisms and the ability of the templates to enhance or decrease the nucleation probability. In this study, we demonstrate that the ability of the seeds to promote the formation of precursors (pre-ordered liquid regions) with enhanced bond-orientational order and favourable structural hallmarks determines to a large extend the nucleating efficiency at the template and polymorph selection(see Fig.~\ref{fig:Ni-hetnuc}). In view of these findings and previous studies that have reported precursor-mediated crystallization mechanisms in a variety of systems, we have proposed a new perspective on  `precursor-mediated' heterogeneous nucleation, where the hidden structural features of the liquid become hallmarks of the nucleating ability of templates.  

Another very recent study of ice nucleation at surfaces~\cite{davies_AccuratePrediction_2022} discusses what the structural features of the liquid can disclose about the nucleating ability of substrates. Using deep learning methods, Davies {\it et al.}~\cite{davies_AccuratePrediction_2022} predicted the nucleation ability of a broad range of substrates from images of  liquid water layers at the interfaces obtained from MD simulations. These findings not only confirm a strong interconnection between structural heterogeneity in the liquid and crystal nucleation, but provide novel ideas that could be the base of other artificial intelligence studies to screen and classify substrates, allowing fast predictions of the nucleating ability (see also discussion in section~\ref{sec:future}). 

Compared to structural heterogeneity in the liquid, our understanding of the connection between dynamical heterogeneity at interfaces and crystallization remains sparse. 
Gasparotto {\it et al.}~\cite{D2NR00387B} showed that interfacial water films and nano-droplets can perturb the dynamics of water forming persistent, spatially extended dynamical domains with an average mobility that varies as a function of the distance from the interface. The dynamical response was observed to vary with the definition of the interface.  These findings hint that liquid mobility domains at interfaces could also signal the nucleating ability of templates and the enhancement or frustration of the nucleation probability. 

The connection between dynamical and structural heterogeneity in the liquid and crystallization suggests that classical and longstanding views of heterogeneous nucleation during crystallization  need to be revised and reformulated. But, most importantly, our current understanding of the fundamental link between liquid heterogeneity and crystallization paves a way to novel scientific and technological advances that may allow us to derive predictive rules for the control of template-driven crystallization and polymorph selection, as well as the fast screening and classification of nucleating agents with machine learning techniques.

\section{Future directions}
\label{sec:future}

With the help of molecular simulations, tremendous progress has been achieved in understanding the microscopic mechanisms of nucleation processes.  One of the key questions that one seeks to answer is:  is it possible to predict the nucleating ability and polymorph selectivity of a nucleating agent, impurity, or surface?  Likewise, it would be desirable to know what kind of properties a nucleating agent needs to possess to demonstrate a certain nucleating ability or polymorph selectivity.  
As in many other fields, machine learning (ML) approaches have become increasingly popular to tackle a variety of challenges in molecular simulations.

\subsection{Machine learning nucleating abilities}

Two recent approaches have explored the application of supervised ML to predict the ice nucleation ability (represented by the nucleation temperature) of various types of substrates~\cite{fitzner_PredictingHeterogeneous_2020,davies_AccuratePrediction_2022}.  To train an ML model, the corresponding data as well as a suitable representation of the data are required.  The data in these studies were produced by molecular dynamics simulations for a large number of diverse substrate where the nucleating temperature was estimated from cooling ramps.  The data representation was quite different in the two cases.  In the first approach~\cite{fitzner_PredictingHeterogeneous_2020}, a large number of descriptors was computed as input to the ML model.  These descriptors characterised the properties of both the substrate and the liquid.
A cluster based feature selection identified the most relevant descriptors that were then used in the ML model to predict the nucleation temperature.  Interestingly, of the four most important descriptors, two were correlated with the properties of the liquid and two with those of the substrate.  
In the second approach~\cite{davies_AccuratePrediction_2022}, the ML model was only trained on the properties of the liquid.  Instead of computing descriptors, the density of the first layer of water in contact with the substrate was projected into two dimensions and the corresponding image was used as input to the ML model.  This is quite interesting as only the effect of the substrate on the structure of the liquid is included but not the properties of the substrate itself, corroborating the perspective on heterogeneous nucleation discussed in section~\ref{sec:heteronuc}: efficient nucleating agents enhance the formation of favourable precursors in the liquid.  
Such an ML model could then be employed to predict the nucleating ability of any agents simply by analysing the structure of the liquid in its vicinity.
It is, however, still unclear how to include changes in the external conditions, such as changes in the cooling rate or pressure.  It also remains to be seen how well such an ML approach works for other systems.  The bottleneck here is certainly the acquisition of reliable and sufficient data to train the ML model.
Specifically, the estimation of nucleating temperatures from molecular simulations hinges on the accuracy of the employed force field as well as the validity of the  simulation approach (e.g., choice of cooling rates, system size, etc.).  Training instead on the properties of the supercooled liquid and interfaces may provide an alternative that is less limited in this respect.

ML models could also be useful to predict more fundamental properties.  
A number of supervised and unsupervised ML approaches have been suggested for structure classification~\cite{geiger_NeuralNetworks_2013,boattini_UnsupervisedLearning_2019,defever_GeneralizedDeep_2019,fulford_DeepIceDeep_2019,kim_GCIceNetGraph_2020,reinhart_UnsupervisedLearning_2021,wang_DescriptorfreeUnsupervised_2022}, which may also improve the analysis of local structural motifs, in particular in the early stages nucleation, and, consequentially, yield new insights into the role of structural fluctuations and other non-classical effects~\cite{leoni_NonclassicalNucleation_2021}. For dynamical properties, much less has been proposed so far.
In a recent study~\cite{boattini_AveragingLocal_2021}, an ML model was trained to predict the dynamical propensity in the supercooled liquid of a glassy system.  The featurization of the local environment around each atom was a central ingredient, establishing a correlation between local structural motifs and dynamical properties.  In the context of nucleation, such an approach could be helpful to recognise regions of low mobility that are correlated with domains having a high probability for the formation of structural precursors and the subsequent emergence of crystalline clusters.

\subsection{Controlling nucleation processes}

If the dynamical and structural properties of the supercooled liquid can be modified in a well-defined manner, this could provide novel technological routes towards the targeted nucleation of specific polymorphs and screening the nucleating ability of substrates. Apart from nucleating agents, such as substrates or other impurities~\cite{Sosso2016,Cacciuto2004,Villeneuve_2005,Jungblut2013, Jungblut_2011, Jungblut2016b,PhysRevLett.100.108302}, 
targeted modifications of the nucleation events can  be triggered experimentally by modifying the liquid structure with external stimuli such as lasers, electric fields, or elongational flows. Advances in experimental techniques have made it possible to manipulate colloidal suspensions and create pattern templates using, e.g., optical tweezers~\cite{doi:10.1063/1.1522397,van_blaaderen_template-directed_1997,doi:10.1063/1.1784559,C9SM01297D,C0SM01219J}. In these experiments, templates are created by fixing the positions of atoms in the liquid to investigate the impact on the polymorphs that crystallize and the nucleation rates. Pre-ordering of a liquid by elongational flow is of extreme practical importance for polymer crystallization. Pressure quench and shear flows, in combination with small- and wide-angle X-Ray scattering for detection, have been employed recently to influence the stability of precursors during polymer crystallization~\cite{doi:10.1021/ma2027325}. Dynamic light scattering, Brownian microscopy and atomic force microscopy are often used to detect precursor clusters in protein crystallization and could be used to screen the nucleating ability of these systems~\cite{Vekilova2014}. Ultrashort pulse laser irradiation at a high repetition rate is nowadays able to recrystallize metallic glasses by modifying the amorphous structure of the glass and inducing the appearance of embedded textured nanocrystals~\cite{doi:10.1021/acs.cgd.8b01802,ANTONOWICZ2021161437}. Ultrasound waves, electric and magnetic fields are often used to control and influence ice nucleation, but the use of electric fields has primarily been employed for protein crystallization studies~\cite{doi:10.1021/jz201113m,DALVIISFAHAN2017222,
C9CE00755E}. Such experiments have great implications for our understanding of crystallization processes and are clear examples of applications where targeted structural and dynamical modifications of the liquid could be employed to influence the nucleation probability and polymorphs selectivity. 

To conclude, our current perspective on the microscopic processes governing nucleation mechanisms suggests that the key in the design of nucleating agents that promote/inhibit nucleation and exhibit specific polymorph selectivity  is the control of the dynamical and structural properties of the liquid.

\enlargethispage{20pt}

\vskip6pt

\aucontribute{JR and GDL conceptualised the manuscript, wrote the original draft, and performed the final review and editing.}

\competing{The authors declare that they have no competing interests.}

\funding{JR acknowledges financial support from the Deut\-sche Forschungsgemeinschaft (DFG) through the Heisenberg Programme project 428315600.
GDL acknowledges support from Conacyt-Mexico through fellowship Ref. No. 220644.
}



\begin{thebibliography}{10}
\expandafter\ifx\csname urlstyle\endcsname\relax
  \providecommand{\doi}[1]{(doi:\discretionary{}{}{}#1)}\else
  \providecommand{\doi}[1]{\begingroup \urlstyle{rm}
  (\href{http://dx.doi.org/#1}{doi:\discretionary{}{}{} \nolinkurl{#1}})
  \endgroup }\fi

\bibitem{Sosso2016}
Sosso GC, Chen J, Cox SJ, Fitzner M, Pedevilla P, Zen A, Michaelides A. 2016
  Crystal nucleation in liquids: Open questions and future challenges in
  molecular dynamics simulations.
\newblock \emph{Chem. Rev.} \textbf{116}, 7078--7116.
\newblock \doi{10.1021/acs.chemrev.5b00744}

\bibitem{Jungblut2016}
Jungblut S, Dellago C. 2016 Pathways to self-organization: Crystallization via
  nucleation and growth.
\newblock \emph{Eur. Phys. J. E} \textbf{39}, 77.
\newblock \doi{10.1021/acs.chemrev.5b00744}

\bibitem{Blow2021}
Blow KE, Quigley D, Sosso GC. 2021 The seven deadly sins: {{When}} computing
  crystal nucleation rates, the devil is in the details.
\newblock \emph{J. Chem. Phys.} \textbf{155}, 040901.
\newblock \doi{10.1063/5.0055248}

\bibitem{Becker1935}
Becker R, D\"{o}ring W. 1935 {Kinetische Behandlung der Keimbildung in
  {\"u}bers{\"a}ttigten D{\"a}mpfen}.
\newblock \emph{Ann. Phys.} \textbf{416}, 719--752.
\newblock \doi{10.1002/andp.19354160806}

\bibitem{Binder1987}
Binder K. 1987 {Theory of first-order phase transitions}.
\newblock \emph{Rep. Prog. Phys.} \textbf{50}, 783--859.
\newblock \doi{10.1088/0034-4885/50/7/001}

\bibitem{Gebauer2014}
Gebauer D, Kellermeier M, Gale JD, Bergström L, Cölfen H. 2014 Pre-nucleation
  clusters as solute precursors in crystallisation.
\newblock \emph{Chem. Soc. Rev.} \textbf{43}, 2348--2371.
\newblock \doi{10.1039/C3CS60451A}

\bibitem{tenWolde1997}
ten Wolde PR, Frenkel D. 1997 Enhancement of protein crystal nucleation by
  critical density fluctuations.
\newblock \emph{Science} \textbf{277}, 1975--1978.
\newblock \doi{10.1126/science.277.5334.1975}

\bibitem{tenWolde1999}
{ten Wolde} PR, Frenkel D. 1999 Homogeneous nucleation and the {{Ostwald}} step
  rule.
\newblock \emph{Phys. Chem. Chem. Phys.} \textbf{1}, 2191--2196.
\newblock \doi{10.1039/a809346f}

\bibitem{Zhang2007}
Zhang TH, Liu XY. 2007 {How Does a Transient Amorphous Precursor Template
  Crystallization}.
\newblock \emph{J. Am. Chem. Soc.} \textbf{129}, 13520.
\newblock \doi{10.1021/ja073598k}

\bibitem{prestipino_SystematicImprovement_2012}
Prestipino S, Laio A, Tosatti E. 2012 Systematic {{Improvement}} of {{Classical
  Nucleation Theory}}.
\newblock \emph{Phys. Rev. Lett.} \textbf{108}, 225701.
\newblock \doi{10.1103/PhysRevLett.108.225701}

\bibitem{Russo2012}
Russo J, Tanaka H. 2012 {The microscopic pathway to crystallization in
  supercooled liquids.}
\newblock \emph{Sci. Rep.} \textbf{2}, 505--512.
\newblock \doi{10.1038/srep00505}

\bibitem{Sear2012}
Sear RP. 2012 The non-classical nucleation of crystals: Microscopic mechanisms
  and applications to molecular crystals, ice and calcium carbonate.
\newblock \emph{Int. Mater. Rev.} \textbf{57}, 328.
\newblock \doi{10.1179/1743280411Y.0000000015}

\bibitem{DLeines2017}
D\'{i}az~Leines G, Drautz R, Rogal J. 2017 Atomistic insight into the
  non-classical nucleation mechanism during solidification in {Ni}.
\newblock \emph{J. Chem. Phys.} \textbf{146}, 154702.
\newblock \doi{10.1063/1.4980082}

\bibitem{Lechner2010}
Lechner W, Rogal J, Juraszek J, Ensing B, Bolhuis PG. 2010 Nonlinear reaction
  coordinate analysis in the reweighted path ensemble.
\newblock \emph{J. Chem. Phys.} \textbf{133}, 174110.
\newblock \doi{10.1063/1.3491818}

\bibitem{fitzner_PredictingHeterogeneous_2020}
Fitzner M, Pedevilla P, Michaelides A. 2020 Predicting heterogeneous ice
  nucleation with a data-driven approach.
\newblock \emph{Nat. Commun.} \textbf{11}, 4777.
\newblock \doi{10.1038/s41467-020-18605-3}

\bibitem{PhysRevLett.108.025502}
T\'oth GI, Tegze G, Pusztai T, Gr\'an\'asy L. 2012 Heterogeneous crystal
  nucleation: The effect of lattice mismatch.
\newblock \emph{Phys. Rev. Lett.} \textbf{108}, 025502.
\newblock \doi{10.1103/PhysRevLett.108.025502}

\bibitem{Jungblut2013}
Jungblut S, Dellago C. 2013 {Crystallization on prestructured seeds}.
\newblock \emph{Phys. Rev. E} \textbf{87}, 012305.
\newblock \doi{10.1103/PhysRevE.87.012305}

\bibitem{Page2009}
Page AJ, Sear RP. 2009 Crystallization controlled by the geometry of a surface.
\newblock \emph{J. Am. Chem. Soc.} \textbf{131}, 17550.
\newblock \doi{doi: 10.1021/ja9085512}

\bibitem{doi:10.1063/1.4961652}
Lupi L, Peters B, Molinero V. 2016 Pre-ordering of interfacial water in the
  pathway of heterogeneous ice nucleation does not lead to a two-step
  crystallization mechanism.
\newblock \emph{J. Chem. Phys.} \textbf{145}, 211910.
\newblock \doi{10.1063/1.4961652}

\bibitem{PhysRevLett.128.166001}
D\'{\i}az~Leines G, Rogal J. 2022 Template-induced precursor formation in
  heterogeneous nucleation: Controlling polymorph selection and nucleation
  efficiency.
\newblock \emph{Phys. Rev. Lett.} \textbf{128}, 166001.
\newblock \doi{10.1103/PhysRevLett.128.166001}

\bibitem{DLeines2018}
D\'{i}az~Leines G, Rogal J. 2018 Maximum likelihood analysis of reaction
  coordinates during solidification in {Ni}.
\newblock \emph{J. Phys. Chem. B} \textbf{122}, 10934--10942.
\newblock \doi{10.1021/acs.jpcb.8b08718}

\bibitem{Piaggi_icenuc}
Piaggi PM, Weis J, Panagiotopoulos AZ, Debenedetti P, Car R. 2022 Homogeneous
  ice nucleation in an ab initio machine-learning model of water.
\newblock \emph{Proc. Natl. Acad. Sci. USA} \textbf{119}, e2207294119.
\newblock \doi{10.1073/pnas.2207294119}

\bibitem{Tanaka2012}
Tanaka H. 2012 Bond orientational order in liquids: Towards a unified
  description of water-like anomalies, liquid-liquid transition, glass
  transition, and crystallization.
\newblock \emph{Eur. Phys. J. E} \textbf{35}, 113.
\newblock \doi{10.1140/epje/i2012-12113-y}

\bibitem{PhysRevX.8.021040}
Russo J, Romano F, Tanaka H. 2018 Glass forming ability in systems with
  competing orderings.
\newblock \emph{Phys. Rev. X} \textbf{8}, 021040.
\newblock \doi{10.1103/PhysRevX.8.021040}

\bibitem{fitzner_IceBorn_2019}
Fitzner M, Sosso GC, Cox SJ, Michaelides A. 2019 Ice is born in low-mobility
  regions of supercooled liquid water.
\newblock \emph{Proc. Natl. Acad. Sci. U.S.A.} \textbf{116}, 2009--2014.
\newblock \doi{10.1073/pnas.1817135116}

\bibitem{Zhang2019}
Zhang Q, Wang J, Tang S, Wang Y, Li J, Zhou W, Wang Z. 2019 Molecular dynamics
  investigation of the local structure in iron melts and its role in crystal
  nucleation during rapid solidification.
\newblock \emph{Phys. Chem. Chem. Phys.} \textbf{21}, 4122--4135.
\newblock \doi{10.1039/C8CP05654D}

\bibitem{doi:10.1063/5.0017575}
Menon S, D\'{i}az~Leines G, Drautz R, Rogal J. 2020 Role of pre-ordered liquid
  in the selection mechanism of crystal polymorphs during nucleation.
\newblock \emph{J. Chem. Phys.} \textbf{153}, 104508.
\newblock \doi{10.1063/5.0017575}

\bibitem{Hu2022}
Hu YC, Tanaka H. 2022 Revealing the role of liquid preordering in
  crystallisation of supercooled liquids.
\newblock \emph{Nat. Commun.} \textbf{13}, 4519.
\newblock \doi{10.1038/s41467-022-32241-z}

\bibitem{PhysRevLett.105.025701}
Schilling T, Sch\"ope HJ, Oettel M, Opletal G, Snook I. 2010 Precursor-mediated
  crystallization process in suspensions of hard spheres.
\newblock \emph{Phys. Rev. Lett.} \textbf{105}, 025701.
\newblock \doi{10.1103/PhysRevLett.105.025701}

\bibitem{PhysRevLett.96.175701}
Sch\"ope HJ, Bryant G, van Megen W. 2006 Two-step crystallization kinetics in
  colloidal hard-sphere systems.
\newblock \emph{Phys. Rev. Lett.} \textbf{96}, 175701.
\newblock \doi{10.1103/PhysRevLett.96.175701}

\bibitem{Lechner2011a}
Lechner W, Dellago C, Bolhuis PG. 2011 {Role of the prestructured surface cloud
  in crystal nucleation}.
\newblock \emph{Phys. Rev. Lett.} \textbf{106}, 085701.
\newblock \doi{10.1103/PhysRevLett.106.085701}

\bibitem{Tan2014}
Tan P, Xu N, Xu L. 2014 Visualizing kinetic pathways of homogeneous nucleation
  in colloidal crystallization.
\newblock \emph{Nat. Phys..} \textbf{10}, 73.
\newblock \doi{doi:10.1038/nphys2817}

\bibitem{davies_AccuratePrediction_2022}
Davies MB, Fitzner M, Michaelides A. 2022 Accurate prediction of ice nucleation
  from room temperature water.
\newblock \emph{Proc. Natl. Acad. Sci. U.S.A.} \textbf{119}, e2205347119.
\newblock \doi{10.1073/pnas.2205347119}

\bibitem{PhysRevLett.96.046102}
Lutsko JF, Nicolis G. 2006 Theoretical evidence for a dense fluid precursor to
  crystallization.
\newblock \emph{Phys. Rev. Lett.} \textbf{96}, 046102.
\newblock \doi{10.1103/PhysRevLett.96.046102}

\bibitem{doi:10.1021/cg049977w}
Vekilov PG. 2004 Dense liquid precursor for the nucleation of ordered solid
  phases from solution.
\newblock \emph{Crystal Growth \& Design} \textbf{4}, 671--685.
\newblock \doi{10.1021/cg049977w}

\bibitem{diazleines_InterplayStructural_2022}
D{\'i}az~Leines G, Michaelides A, Rogal J. 2022 Interplay of structural and
  dynamical heterogeneity in the nucleation mechanism in nickel.
\newblock \emph{Faraday Discuss.} \textbf{235}, 406--415.
\newblock \doi{10.1039/D1FD00099C}

\bibitem{PhysRevLett.102.198302}
Savage JR, Dinsmore AD. 2009 Experimental evidence for two-step nucleation in
  colloidal crystallization.
\newblock \emph{Phys. Rev. Lett.} \textbf{102}, 198302.
\newblock \doi{10.1103/PhysRevLett.102.198302}

\bibitem{doi:10.1063/1.1992475}
O’Malley B, Snook I. 2005 Structure of hard-sphere fluid and precursor
  structures to crystallization.
\newblock \emph{The Journal of Chemical Physics} \textbf{123}, 054511.
\newblock \doi{10.1063/1.1992475}

\bibitem{Kawasaki2011}
Kawasaki T, Tanaka H. 2010 {Formation of a crystal nucleus from liquid}.
\newblock \emph{Proc. Natl. Acad. Sci. USA} \textbf{107}, 14036–--14041.
\newblock \doi{10.1073/pnas.1001040107}

\bibitem{Dellago2002}
Dellago C, Bolhuis P, Geissler PL. 2002 {Transition Path Sampling}.
\newblock \emph{Adv. Chem. Phys.} \textbf{123}, 1--78.
\newblock \doi{10.1002/0471231509.ch1}

\bibitem{VanErp2005}
van Erp TS, Bolhuis PG. 2005 {Elaborating transition interface sampling
  methods}.
\newblock \emph{J. Comp. Phys.} \textbf{205}, 157--181.
\newblock \doi{10.1016/j.jcp.2004.11.003}

\bibitem{Rogal2010}
Rogal J, Lechner W, Juraszek J, Ensing B, Bolhuis PG. 2010 {The reweighted path
  ensemble}.
\newblock \emph{J. Chem. Phys.} \textbf{133}, 174109.
\newblock \doi{10.1063/1.3491817}

\bibitem{Lechner2008}
Lechner W, Dellago C. 2008 Accurate determination of crystal structures based
  on averaged local bond order parameters.
\newblock \emph{J. Chem. Phys.} \textbf{129}, 114707.
\newblock \doi{10.1063/1.2977970}

\bibitem{berthier_TheoreticalPerspective_2011}
Berthier L, Biroli G. 2011 Theoretical perspective on the glass transition and
  amorphous materials.
\newblock \emph{Rev. Mod. Phys.} \textbf{83}, 587--645.
\newblock \doi{10.1103/RevModPhys.83.587}

\bibitem{toninelli_DynamicalSusceptibility_2005a}
Toninelli C, Wyart M, Berthier L, Biroli G, Bouchaud JP. 2005 Dynamical
  susceptibility of glass formers: {{Contrasting}} the predictions of
  theoretical scenarios.
\newblock \emph{Phys. Rev. E} \textbf{71}, 041505.
\newblock \doi{10.1103/PhysRevE.71.041505}

\bibitem{sosso_DynamicalHeterogeneity_2014}
Sosso GC, Colombo J, Behler J, Del~Gado E, Bernasconi M. 2014 Dynamical
  {{Heterogeneity}} in the {{Supercooled Liquid State}} of the {{Phase Change
  Material GeTe}}.
\newblock \emph{J. Phys. Chem. B} \textbf{118}, 13621--13628.
\newblock \doi{10.1021/jp507361f}

\bibitem{widmer-cooper_StudyCollective_2007}
{Widmer-Cooper} A, Harrowell P. 2007 On the study of collective dynamics in
  supercooled liquids through the statistics of the isoconfigurational
  ensemble.
\newblock \emph{J. Chem. Phys.} \textbf{126}, 154503.
\newblock \doi{10.1063/1.2719192}

\bibitem{colombo_MicroscopicPicture_2013}
Colombo J, {Widmer-Cooper} A, Del~Gado E. 2013 Microscopic {{Picture}} of
  {{Cooperative Processes}} in {{Restructuring Gel Networks}}.
\newblock \emph{Phys. Rev. Lett.} \textbf{110}, 198301.
\newblock \doi{10.1103/PhysRevLett.110.198301}

\bibitem{Turnbull1952}
Turnbull D, Vonnegut B. 1952 Nucleation catalysis.
\newblock \emph{Ind. Eng. Chem.} \textbf{44}, 1292.
\newblock \doi{10.1021/ie50510a031}

\bibitem{fitzner_precrfluc_2017}
Fitzner M, Sosso G, Pietrucci F, Pipolo S, Michaelides A. 2017 Pre-critical
  fluctuations and what they disclose about heterogeneous crystal nucleation.
\newblock \emph{Nat. Commun.} \textbf{8}, 2257.
\newblock \doi{10.1038/s41467-017-02300-x}

\bibitem{D2NR00387B}
Gasparotto P, Fitzner M, Cox SJ, Sosso GC, Michaelides A. 2022 How do
  interfaces alter the dynamics of supercooled water?
\newblock \emph{Nanoscale} \textbf{14}, 4254--4262.
\newblock \doi{10.1039/D2NR00387B}

\bibitem{geiger_NeuralNetworks_2013}
Geiger P, Dellago C. 2013 Neural networks for local structure detection in
  polymorphic systems.
\newblock \emph{J. Chem. Phys.} \textbf{139}, 164105.
\newblock \doi{10.1063/1.4825111}

\bibitem{boattini_UnsupervisedLearning_2019}
Boattini E, Dijkstra M, Filion L. 2019 Unsupervised learning for local
  structure detection in colloidal systems.
\newblock \emph{J. Chem. Phys.} \textbf{151}, 154901.
\newblock \doi{10.1063/1.5118867}

\bibitem{defever_GeneralizedDeep_2019}
DeFever RS, Targonski C, Hall SW, Smith MC, Sarupria S. 2019 A generalized deep
  learning approach for local structure identification in molecular
  simulations.
\newblock \emph{Chem. Sci.} \textbf{10}, 7503--7515.
\newblock \doi{10.1039/C9SC02097G}

\bibitem{fulford_DeepIceDeep_2019}
Fulford M, Salvalaglio M, Molteni C. 2019 {{DeepIce}}: {{A Deep Neural Network
  Approach To Identify Ice}} and {{Water Molecules}}.
\newblock \emph{J. Chem. Inf. Model.} \textbf{59}, 2141--2149.
\newblock \doi{10.1021/acs.jcim.9b00005}

\bibitem{kim_GCIceNetGraph_2020}
Kim Q, Ko JH, Kim S, Jhe W. 2020 {{GCIceNet}}: A graph convolutional network
  for accurate classification of water phases.
\newblock \emph{Phys. Chem. Chem. Phys.} \textbf{22}, 26340--26350.
\newblock \doi{10.1039/D0CP03456H}

\bibitem{reinhart_UnsupervisedLearning_2021}
Reinhart WF. 2021 Unsupervised learning of atomic environments from simple
  features.
\newblock \emph{Comput. Mater. Sci.} \textbf{196}, 110511.
\newblock \doi{10.1016/j.commatsci.2021.110511}

\bibitem{wang_DescriptorfreeUnsupervised_2022}
Wang Y, Deng W, Huang Z, Li S. 2022 Descriptor-free unsupervised learning
  method for local structure identification in particle packings.
\newblock \emph{J. Chem. Phys.} \textbf{156}, 154504.
\newblock \doi{10.1063/5.0088056}

\bibitem{leoni_NonclassicalNucleation_2021}
Leoni F, Russo J. 2021 Nonclassical {{Nucleation Pathways}} in
  {{Stacking-Disordered Crystals}}.
\newblock \emph{Phys. Rev. X} \textbf{11}, 031006.
\newblock \doi{10.1103/PhysRevX.11.031006}

\bibitem{boattini_AveragingLocal_2021}
Boattini E, Smallenburg F, Filion L. 2021 Averaging {{Local Structure}} to
  {{Predict}} the {{Dynamic Propensity}} in {{Supercooled Liquids}}.
\newblock \emph{Phys. Rev. Lett.} \textbf{127}, 088007.
\newblock \doi{10.1103/PhysRevLett.127.088007}

\bibitem{Cacciuto2004}
Cacciuto A, Auer S, Frenkel D. 2004 Onset of heterogeneous crystal nucleation
  in colloidal suspensions.
\newblock \emph{Nature} \textbf{428}, 404.
\newblock \doi{10.1038/nature02397}

\bibitem{Villeneuve_2005}
de~Villeneuve VWA, Verboekend D, Dullens RPA, Aarts DGAL, Kegel WK,
  Lekkerkerker HNW. 2005 Hard sphere crystal nucleation and growth near large
  spherical impurities.
\newblock \emph{J. Phys.: Condens. Matter} \textbf{17}, S3371.
\newblock \doi{10.1088/0953-8984/17/45/024}

\bibitem{Jungblut_2011}
Jungblut S, Dellago C. 2011 Heterogeneous crystallization on tiny clusters.
\newblock \emph{Eur. Phys. Lett.} \textbf{96}, 56006.
\newblock \doi{10.1209/0295-5075/96/56006}

\bibitem{Jungblut2016b}
Jungblut S, Dellago C. 2016 Heterogeneous crystallization on pairs of
  pre-structured seeds.
\newblock \emph{J. Phys. Chem. B} \textbf{120}, 9230.
\newblock \doi{10.1021/acs.jpcb.6b06510}

\bibitem{PhysRevLett.100.108302}
van Teeffelen S, Likos CN, L\"owen H. 2008 Colloidal crystal growth at
  externally imposed nucleation clusters.
\newblock \emph{Phys. Rev. Lett.} \textbf{100}, 108302.
\newblock \doi{10.1103/PhysRevLett.100.108302}

\bibitem{doi:10.1063/1.1522397}
Hoogenboom JP, Derks D, Vergeer P, van Blaaderen A. 2002 Stacking faults in
  colloidal crystals grown by sedimentation.
\newblock \emph{J. Chem. Phys.} \textbf{117}, 11320.
\newblock \doi{10.1063/1.1522397}

\bibitem{van_blaaderen_template-directed_1997}
van Blaaderen A, Ruel R, Wiltzius P. 1997 Template-directed colloidal
  crystallization.
\newblock \emph{Nature} \textbf{385}, 321.
\newblock \doi{10.1038/385321a0}

\bibitem{doi:10.1063/1.1784559}
Vossen DLJ, van~der Horst A, Dogterom M, van Blaaderen A. 2004 Optical tweezers
  and confocal microscopy for simultaneous three-dimensional manipulation and
  imaging in concentrated colloidal dispersions.
\newblock \emph{Rev. Sci. Instrum.} \textbf{75}, 2960.
\newblock \doi{10.1063/1.1784559}

\bibitem{C9SM01297D}
Walton F, Wynne K. 2019 Using optical tweezing to control phase separation and
  nucleation near a liquid–liquid critical point.
\newblock \emph{Soft Matter} \textbf{15}, 8279--8289.
\newblock \doi{10.1039/C9SM01297D}

\bibitem{C0SM01219J}
Hermes M, Vermolen ECM, Leunissen ME, Vossen DLJ, van Oostrum PDJ, Dijkstra M,
  van Blaaderen A. 2011 Nucleation of colloidal crystals on configurable seed
  structures.
\newblock \emph{Soft Matter} \textbf{7}, 4623.
\newblock \doi{10.1039/C0SM01219J}

\bibitem{doi:10.1021/ma2027325}
Ma Z, Balzano L, Peters GWM. 2012 Pressure quench of flow-induced
  crystallization precursors.
\newblock \emph{Macromolecules} \textbf{45}, 4216--4224.
\newblock \doi{10.1021/ma2027325}

\bibitem{Vekilova2014}
Vekilova PG, Vorontsovaa M. 2014 Nucleation precursors in protein
  crystallization.
\newblock \emph{Acta Crystallogr. F: Struct. Biol. Commun.} \textbf{1},
  271–282.
\newblock \doi{10.1107/S2053230X14002386}

\bibitem{doi:10.1021/acs.cgd.8b01802}
Cao J, Lancry M, Brisset F, Mazerolles L, Saint-Martin R, Poumellec B. 2019
  Femtosecond laser-induced crystallization in glasses: Growth dynamics for
  orientable nanostructure and nanocrystallization.
\newblock \emph{Cryst. Growth Des.} \textbf{19}, 2189--2205.
\newblock \doi{10.1021/acs.cgd.8b01802}

\bibitem{ANTONOWICZ2021161437}
Antonowicz J, Zalden P, Sokolowski-Tinten K, Georgarakis K, Minikayev R,
  Pietnoczka A, Bertram F, Chaika M, Chojnacki M, D\l{}u$\dot{\text{z}}$ewski
  P, Fronc K, Greer A, Jastrz\c{e}bski C, Klinger D, Lemke C, Magnussen O,
  Murphy B, Perumal K, Ruett U, Warias J, Sobierajski R. 2021 Devitrification
  of thin film {Cu–Zr} metallic glass via ultrashort pulsed laser annealing.
\newblock \emph{J. Alloys Compd.} \textbf{887}, 161437.
\newblock \doi{10.1016/j.jallcom.2021.161437}

\bibitem{doi:10.1021/jz201113m}
Yan JY, Patey GN. 2011 Heterogeneous ice nucleation induced by electric fields.
\newblock \emph{J. Phys. Chem. Lett.} \textbf{2}, 2555--2559.
\newblock \doi{10.1021/jz201113m}

\bibitem{DALVIISFAHAN2017222}
Dalvi-Isfahan M, Hamdami M, Xanthakis E, Le-Bail A. 2017 Review on the control
  of ice nucleation by ultrasound waves, electric and magnetic fields.
\newblock \emph{J. Food Eng.} \textbf{195}, 222--234.
\newblock \doi{10.1016/j.jfoodeng.2016.10.001}

\bibitem{C9CE00755E}
Alexander LF, Radacsi N. 2019 Application of electric fields for controlling
  crystallization.
\newblock \emph{CrystEngComm} \textbf{21}, 5014--5031.
\newblock \doi{10.1039/C9CE00755E}

\end{thebibliography}
\end{document}